**Modulation of N- to C-terminal interactions enhances protein stability.**

Mahanta P[a], Bhardwaj A[b,c], Kumar K[b], Reddy V S[b], Ramakumar S[a]∗

[a]Department of Physics, Indian Institute of Science, Bangalore-560012; [b]Plant Transformation Group, International Centre for Genetic Engineering and Biotechnology, New Delhi-110067; [c]Present address, Department of Pathology, Kimmel Centre for Biology and Medicine at the Skirball Institute, New York University School of Medicine, New York, 10016, USA

* Corresponding author: ramak@physics.iisc.ernet.in

**Abstract**

Although, several factors have been attributed to thermostability, the stabilization strategies used by proteins are still enigmatic. Studies on recombinant xylanase which has the ubiquitous $(\beta/\alpha)_8$ TIM (Triosephosphate isomerase) barrel fold showed that, just a single extreme N-terminus mutation (V1L) markedly enhanced the thermostability by 5 °C without loss of catalytic activity whereas another mutation, V1A at the same position decreased the stability by 2 °C. Based on computational analysis of their crystal structures including residue interaction network, we established a link between N- to C-terminal contacts and protein stability. We demonstrate that augmenting of N- to C-terminal non-covalent interactions is associated with the enhancement of protein stability. We propose that the strategy of mutations at the termini could be exploited with a view to modulate stability without compromising on enzymatic activity, or in general, protein function, in diverse folds where N- and C-termini are in close proximity. Finally, we discuss the implications of our results for the development of therapeutics involving proteins and for designing effective protein engineering strategies.

**Introduction**

Elucidating the molecular mechanisms of protein stability at high temperature continues to attract and fascinate researchers over a broad range of disciplines and has still remained a challenging puzzle. A number of approaches have been employed to develop stable proteins for biotechnological applications[1, 2]. Site directed mutagenesis is an attractive approach to provide valuable insights into the structural features that govern protein thermostability. Locating the target site of mutagenesis for stability-improvement can reduce the screening effort required to find stable mutant(s) by orders of magnitude as compared to random directed evolution methods[3].

Enzyme stability and activity often appear to trade off at the level of individual mutations. For example, while flexibility is required for the catalytic activity of most enzymes, higher thermostability necessitates an increase in the rigidity of the structure. As a result, mutants with increased stability often lose catalytic efficiency[4]. In addition, engineering protein thermostability at the expense of losing enzymatic activity is not a biotechnologically desirable outcome. Generally, industrial processes are performed at high temperature. Therefore, improving the stability of an already stable enzyme could be advantageous for industrial applications. Besides, even a modest increase of stability could lead to >10-fold longer lifetime[5,6].

The N and C terminal regions are often overlooked from the point of view of enhancing protein stability. This may be because the terminal regions of a protein structure are more flexible as compared to interior regions. Further, in a majority of cases, the terminal residues are exposed to solvent with low number of nearest neighbors-residues and hence considered to have little influence on thermostability[7]. Nevertheless, certain experimental and computational studies suggest the importance of protein termini on their structure-stability and functions[8, 9]. An in silico analysis of a set of two-state folding proteins revealed the presence of N-C motif (N- to C-terminal contacts) and suggested its possible role in initial protein folding and native state stability[10]. However, there is hardly any available experimental evidence, which clearly brings out in a focused manner the role and importance of interactions between termini in protein stability and how changes in the terminal regions influence stability.

Here, we investigate how, just a single mutation at the extreme N-terminus affects the structure and interactions to change the thermal stability of a biotechnologically important enzyme xylanase, having the ubiquitous TIM (Triosephosphate isomerase) barrel fold. Based on computational analysis including residue interaction network of crystal structure of recombinant xylanase (RBSX) and its mutants, we established the link between protein stability and N- to C-terminal non-covalent interactions. We demonstrated that augmenting of N-to C-terminal non-covalent interactions is associated with the enhancement of stability of protein in fold-specific manner where N- and C-terminus are in close proximity. We observed that even though the mutation was at the extreme N-terminus of the protein, changes are not confined to the terminal regions and they occur throughout the protein including the terminal regions. We show that the cumulative effect of a network of non-covalent interactions which include N-to C-terminal interactions, modulate the thermal stability of the protein. We propose that the mutagenesis at the termini could be exploited with a view to enhance stability without compromising enzymatic activity or protein function. This may be effective especially in situations where the N- and C-termini come close in three-dimensional (3D) space, thereby enabling long-range interactions (interactions between distantly separated residues in primary sequence), as demonstrated with the example of the TIM barrel fold, and the same can be extrapolated to diverse folds in both globular and membrane proteins. Further, the work may elucidate the underlying mechanism dictating the evolution of functional repertoire of the TIM barrel fold. Finally, we discuss the implications of our results for development of therapeutics involving proteins or for designing effective protein engineering strategies.

**Results**

**Structure based rationalization of protein stability.** BSX, an extra-cellular endo-xylanase is a monomeric $(\beta/\alpha)_8$ TIM barrel fold enzyme composed of 354 amino acid residues[11] (Fig. 1). The TIM barrel fold is a common tertiary fold, occurring in many glycosyl hydrolases and is present in approximately 10% of all enzymes[12]. Biophysical/biochemical analysis of different extreme N-terminus mutants of recombinant BSX (RBSX) in our group showed that, a single amino acid substitution, V1→L (V1L) markedly enhanced the thermostability of RBSX from 70 °C to 75 °C without compromising its catalytic activity and showed higher cooperativity in the thermal unfolding transition[13]. On the other hand, substitution of V1→A (V1A) at the same position decreased the stability of the protein from 70 °C to 68 °C[13]. To understand the structural reasons as to how a seemingly unimportant mutation modulates BSX thermal stability, we solved the structure of RBSX and its mutants (V1L and V1A). A brief summary of crystallization, data collections, structure solutions, and refinement statistics are given in Table 1. Crystal structure comparison of mutants with RBSX shows no significant changes in the overall 3D structure of proteins despite their difference in thermal stabilities. The overall $C_\alpha$ root mean square deviation (RMSD) between RBSX and V1L is 0.393 Å whereas that between RBSX and V1A is 0.265 Å. So the question arises, what may be the mechanism of thermal stabilization/destabilization, considering there is only a minimal change in their overall 3D structures? The location of the mutation lies on an extended loop in the extreme N-terminus region and no dominant interaction exclusively made by this first residue (L1 in V1L and V1 in RBSX) is observed. This observation raises the possibility about the effects of non-covalent interactions network that transmits changes near and far from the site of mutation and changes the overall stability of RBSX. In a folded protein, a network of interactions brings the distal residues in a sequence space to close approach in 3D space. This extensive network of interactions gives proteins, structural flexibility, integrity, and thermal-stability[14-16]. Therefore, we focused on residue contacts and residue interaction network (RIN) in the protein structures to capture this cumulative nature of thermo stabilization/ destabilization and to identify the changes in both local and non-local interactions. The aim of the present work is not to obtain the most stable structure by carrying out all possible mutations at the extreme N-terminus, but to gain structural insights into the modulation of stability caused by mutations in the terminal region.

Although, there was an increase of about 5 °C in the thermostability due to a single Leucine mutation, crystal structure analysis showed that L1 in V1L structure has similar type of interactions (van der Waals) as V1 in RBSX structure. However, because of its greater bulk and better conformational accessibility in comparison to V1 in RBSX, Leucine side chain forms more van der Waals contacts with side chain atoms of R344 in the structure (Fig. 2, Table S1). It is possible that these additional interactions help in maintaining the overall protein stability at high temperature. Further, these additional cohesive contacts made by L1, influenced the relative decrease of solvent accessible surface area (SASA) of R344 by 19.6% in V1L in comparison to R344 in RBSX. The decrease of SASA of R344 is even pronounced (29.8 %) when compared between V1L and V1A structure. On the other hand, the decreased stability of V1A structure may be due to the significant lack of van der Waals contacts by A1 in V1A mutant in comparison to V1 of RBSX (Fig. 2 and Table S1). We then looked into the contribution of non-covalent interaction score[17, 18] (which is proportional to the strength of non-covalent interactions) between the mutated residue and R344. We found a higher non-covalent interaction score between L1 and R344 (2.66) for V1L structure than V1 and R344 (0.063) for RBSX structure and A1 and R344 (0.00) for V1A structure.

**The metrics, $C_\beta$ Contact density, and $C_\beta$ Contact order relate to protein stability.** A marked tendency of side chain atoms of L1 to be in close proximity with side chain atoms of R344 in the folded 3D structure was observed. Despite the introduction of such a bulky amino acid residue (L1), we found that the distance between $C_\beta$-atoms of L1 and R344 in V1L mutant structure is smaller (6.48 Å) than that between the corresponding residues in RBSX structure (7.12Å) and V1A structure (7.21 Å). Thus, it may be seen that the distance between $C_\beta$-atoms is shorter in the more stable mutant than the less stable mutant, suggesting that metrics based on $C_\beta$-atoms could be used to study stability changes in the protein. To look into the extent of packing interactions of $C_\beta$-atoms and to assess the effect of $C_\beta$ contact networks due to mutation, we computed and analyzed two $C_\beta$-based structural metrics, $C_\beta$ contact density ($C_\beta$CD) and $C_\beta$ contact order ($C_\beta$CO). $C_\beta$CD is indicative of close packing of $C_\beta$-atoms in the 3D space whereas $C_\beta$CO values depend on degree of long-range interactions in terms of the average magnitudes of residue separation in the primary sequence between the pairs of contacting residues (contribution to $C_\beta$CO will be large, by pairs of residues that have large separation in the primary sequence, see Methods). The analysis of $C_\beta$CD showed that V1L has a higher value of $C_\beta$CD (7.17) than RBSX (7.14) and V1A (7.09) (Table 2). A comparison of $C_\beta$ contacts between V1L and V1A structures reveals that there are 3167 (95.9%) common $C_\beta$ contacts between these two structures, whereas 76 (2.3%) contacts are unique to V1L and 59 (1.8%) are unique to V1A. We observed a substantial difference of long-range $C_\beta$ contacts between V1L and V1A structures. Interestingly, considerable changes are noticed in the N-to C-terminal $C_\beta$ contacts (Fig. S1).

In addition, comparison of $C_\beta CO$ between RBSX and its mutants reveals that $C_\beta CO$ correlates well with the observed changes in RBSX thermostability. V1L has higher $C_\beta CO$ (26.82) followed by RBSX (26.56) and V1A (26.33) (Table 2). $C_\beta CO$ can capture both local and non-local contacts in the 3D structure. To know the influence of each type of contacts in overall value of $C_\beta CO$, we computed $C_\beta CO$ using contacts that are unique to a given structure, e.g. that are only present in one structure and absent in another structure and vice-versa to obtain a better picture about the structural rearrangement of $C_\beta$ contacts upon mutation. Considering only those unique contacts, we observed a much higher difference of $C_\beta CO$ (75.39/53.8) between V1L and V1A structures (Table 2). This result indicates that V1L mutation affects the spatial arrangements of $C_\beta$ contacts via both local and non-local non-covalent interactions and non-local $C_\beta$ contacts are majorly affected. This result is further supported by the higher values of long-range $C_\beta CO$ (LR$C_\beta CO$) for V1L than RBSX and V1A structures, evaluated at different range of residue separations (10, 30, and 50) (Table S2). Taken together, these observations collectively suggest that the substantial increase of V1L thermostability could be a result of better non-local $C_\beta$ contact network that trigger a cascading effect of intra-molecular interactions network throughout the structure.

**Contacts with N-C terminal regions play a role in the protein stability.** We further examined the $C_\beta CO$ at a local, residue level to look at the regions contributing to $C_\beta CO$ by computing residue wise $C_\beta$ contact order (RW$C_\beta CO$). RW$C_\beta CO$ value primarily reflects the extent of long-range contacts in terms of residue separation between contacting residues in a protein structure whereas its properties are based on the residue level. It is apparent that higher values of RWC$_\beta$CO belong to the termini of the protein (Fig. S2). This should be due to the occurrence of a number of contacts between N- and C-terminal regions, and as they are in close proximity in their 3D structure. We then turned our attention to compute ΔRWC$_\beta$CO (RWC$_\beta$CO$_L$-RWC$_\beta$CO$_A$), the difference of RWC$_\beta$CO value between V1L and V1A to look into the measurable changes in the region of protein structure due to mutation and plotted along the primary sequence. Notably, it is observed that there is a net increase of RWC$_\beta$CO in the terminal regions, although small differences are present that mostly correspond to the secondary structure elements (Fig. 3). This is one piece of evidence that terminal regions are majorly affected in comparison to the other parts of the protein structure due to mutation and play an important role in the overall stability of RBSX through long-range interactions.

To assess the contribution of chain termini in RWC$_\beta$CO values and relate them to RBSX stability, we considered in more detail as to how the terminal regions differ in terms of atom-atom contacts. We used a distance cut off 5Å (the higher cutoff for attractive London-van der Waals forces[19]) to capture only effective physical contacts within and between the terminal atoms. Examination of atom-atom contacts between N-terminal segment (residues 1-25; up to second secondary structural elements (SSEs) from N-terminal) and C-terminal segment (residues 319-354 ; up to second SSEs from C-terminal end) for each structure reveals that V1L mutant has a higher value of normalized atom-atom contacts between termini in comparison to V1A (Table S3, Fig. S3). We observed that there is a much larger increase (~11.3%) of contacts between N-terminal segment and C-terminal segment than within the N-terminal segment (~3.4%) in V1L structure compared to V1A structure indicating a cosying up of the terminal regions in the stable mutant.

**Network parameters and protein stability.** Protein structures contain a variety of weak and strong non-covalent interactions that integrate different parts of the structure and interplay of these interactions provides the structural stability to proteins. Here, we analyze this using a network representation of protein structure by generating residue interaction network (RIN) which considers all non-covalent interactions between pairs of interacting residues (see Methods). We also decomposed the network into different sub networks based on the strength of interaction score ($I_s$) between the interacting residues and analyzed their global topology and corresponding network parameters. The analysis reveals that the more stable mutant, V1L has a higher magnitude of most of the network parameters (total number of edges or links (E), edge/node (E/N) ratio; where N is the total number of residues in the protein structure, average nearest neighbors ($<k>$)) than the less stable mutant, V1A (Table S4). The values of these network parameters are very similar to each other. This may be because, all these structures have the same size (354 amino acids), and there is only one amino acid difference in their primary structures. However, it is relevant to compare their network parameters as they show different thermostability scale. We notice that all three parameters (E, E/N and $<k>$) are relatively higher for V1L than RBSX or V1A (Table S5) at all interaction score cut-off ($I_{smin}$s). This may indicate that presence of higher number of interactions (higher E and E/N values) and better connectivity within network ($<k>$) at different $I_{smin}$s are likely to be involved in the structural stability and provides the extra stabilizing force for V1L structure in comparison to RBSX or V1A structures. We can infer that the observed difference in network parameters is a result of combined effect of various subtle interactions (hydrogen bonds, van der Waals, ion pairs etc.) manifested throughout the structure due to a single mutation.

Apart from analyzing the different topological properties, we found that there is a gradual decrease of size of largest strongly connected component (LSCC) with respect to increase of $I_{smin}$ in an edge-weighted RIN of each structure. It is observed that V1L structure has not only higher number of edges and edge/node ratio at different residue interaction sub networks in comparison to V1A, but also larger size of strongly connected components in their network. The cooperative nature of these stabilizing interactions positively influences other interactions as evidenced by largest size of connected component in the more stable mutant structure. Furthermore, a sharp transition in the size of LSCC begins around $I_{smin} = 1$ and lies within a narrow range of $I_{smin}$ (1 to 2), with no major change towards the side of higher interaction cut-off (Fig. S4). This sharp transition in LSCC is attributed to the loss of different non-covalent interactions in the networks as $I_{smin}$ increases, thus quickly generating large number of small clusters. Then we turned our attention to compare the network of largest cluster of amino acid residues at $I_{smin}=1$ for V1L and V1A. The generation of combined comparison network of V1L and V1A based on the superposition alignment of the corresponding 3D structures at $I_{smin} = 1$, results 28 unique residues that correspond to V1L whereas there are 14 unique residues that correspond to V1A (Table S6). In the comparison network, we found 535 number of identical edges that correspond to non-covalent interactions for both V1L and V1A. On the other hand, V1L has considerably higher number of non-identical non-covalent interactions (192 edges) than V1A (125 edges). These numbers reflect that there is a perturbation in the residue interaction networks brought by extreme N-terminus mutation. It is notable that the many of these unique residues are distributed in and around the terminal regions of the protein (Fig. 4). We found that ~25% of these unique residues in V1L and ~14% of unique residues in V1A correspond to the termini. Thus, it may be surmised that terminal residues are important in maintaining the structural stability of these protein structure networks and could be considered for mutation(s) with a view to enhance protein stability. Since the residues, not belongings to the terminal regions are also affected; it would imply that the mutation, apart from directly influencing the interactions between N- to C-terminal regions (Fig. 2), also indirectly influences the interactions involving non-terminal residues (Fig. 4).

**Discussion**

The picture which emerges from our study is that mutation(s) at the extreme N-terminus can be either stabilizing or destabilizing and may be important in a fold such as the TIM barrel fold where the N-terminus and C-terminus come close in 3D space, though separated in sequence. Here, we consider the structural aspects based on crystallographically determined coordinates of RBSX and its mutants, with a view to rationalize the difference in the protein stability due to an extreme N-terminus mutation. Our study establishes that terminal regions should also be considered for mutations for enhancing protein stability in folds where N-terminus and C-terminus come together in 3D structure.

The present findings provide valuable insights into the role of direct non-covalent interactions between N- to C-termini in protein stabilization. Fig. 2 provides an example of such interactions that are enhanced in the more stable mutant. The direct N- to C-terminal contacts in V1L mutant involving L1 shows a clear difference in the degree of packing interactions of the side chain atoms in comparison to V1 in RBSX and A1 in V1A structures (Fig. 2, Table S1). It appears that these additional interactions might be playing a role in tying down the extreme N-terminus during the thermal unfolding at high temperature[12]. Furthermore, we observed an enhancement in the number of overall N- to C-terminal direct contacts in more stable mutant structure (V1L), whereas the absence of many N-to C-terminal contacts could increase local unfolding of the peptide chain at these weak links, and results in a lower unfolding temperature for the V1A mutant (Fig. S3 and Table S3).

Though the mutation is at the extreme N-terminus, the C-terminal region is also affected which shows that changes are not restricted to terminal regions (Fig. 3). We observed that there is structural rearrangement of contacts throughout the structure more so within and between terminal regions. The cooperative nature of these stabilizing interactions indirectly or allosterically propagates to the other parts of the structure and positively influences other interactions as evidenced by network analysis where the largest strongly connected component is bigger in size for the more stable mutant structure (Fig. S4). Thus, it is likely that the increased stability displayed by V1L mutant is a cumulative effect of small changes rather than solely effect of the interactions involving the substituent amino acid. This effect is reminiscent of the concept in economics of 'comedy of the commons' like in property resources[20] applied here to protein stabilization in which a cumulative effect of many contributions leads to a desired outcome, in this case protein stability. In addition, the residues placed at long separation in the primary structure plays an important role in stability of the protein as evidenced by the analysis of $LRC_\beta CO$, $C_\beta CD$, and RINs (Table S3 and Fig. S1). Obviously, N- to C-terminal contacts are the longest-range interactions possible in terms of sequence separation in any given protein (Table S4). These results suggest that the overall increase of long-range interactions (primarily through N- to C-terminal contacts) in V1L structure upon mutation is one of the primary sources of increase in thermal stability. Our results are consistent with the earlier findings that long-range interactions, connecting different parts of the protein structure, have a major role in folding and stabilizing the tertiary structure of the protein[16,21]. However, what is remarkable is that all these structural changes are elicited by just a single mutation at the extreme N-terminus of the protein.

The current study extends our knowledge of the nature of the N-terminal to C-terminal interactions and their role in the stability of a protein. The N and C-termini come together in 3D space and enable stabilization through mutual interactions, a distinct possibility in the TIM barrel fold as demonstrated in the present report. An in silico analysis which was restricted to a set of two-state folding proteins including TIM barrel fold showed the presence of N-C motif (N-terminal to C-terminal contacts) and suggested the possible role in initial protein folding and stability[10]. This view is supported by our findings and crucially provides the experimental evidence in a focused manner through mutations and crystal structure analysis, in reiterating the importance in general of N- to C-terminal interactions on the stability of a protein. Fraying of the terminal regions may make a protein susceptible to unfolding at high temperature. The terminal regions may be stabilized if they interact separately with different parts of the proteins. However, it may be more advantageous if the terminal regions dock with each other and mutually stabilize, thereby reducing susceptibility to unfolding at high temperature.

Our work argues that augmenting N-to C-terminal non-covalent interactions enhances protein stability. Such stabilization presumably insures against unfolding of an already folded protein and it may aid the folding process. Although, it is clearly possible to stabilize proteins with other mechanisms/factors as reported[22,23], we demonstrate that proteins can be stabilized without compromising their biological functions through optimization of N- to C-terminal non-covalent interactions. This apparent stabilization through N- to C-terminal interactions seems to be implicated in the structures of TIM (EC 5.3.1.1) isolated from different organisms. We found higher normalized N- to C-terminal contacts (6.17) for hyperthermophilic TIM from *Thermotoga maritima* (PDB ID: 1B9B, optimum growth temperature (OGT) = 80 °C) than thermophilic TIM (6.15) from *Geobacillus stearothermophilus* (PDB ID: 1BTM, OGT= 65 °C) and mesophilic TIM (5.64) from *Escherichia Coli* (PDB ID: 1TRE, OGT= 37 °C) when compared across monomers, despite their similar 3D structures (Fig. S6). In the case of the TIM barrel fold, the N-C terminal region that can contribute to protein stability belongs to the scaffold region[24,25] and is usually away from the active site region (Fig. 1). This might have contributed to the evolvability of the fold due to which approximately 10% of all enzymes have the TIM barrel fold.

Furthermore, apart from TIM barrel fold proteins, a comparative analysis of an NAD(P)-binding Rossmann-fold domain protein, GAPDH (glyceraldehyde 3-phosphate dehydrogenase) from different bacterial species in which N- and C-terminal are close together reveals higher normalized N- to C-terminal contacts in accordance to their OGT of the organisms. We observed that GAPDH from hyperthermophilic bacterium *Thermotoga maritima* (OGT = 80 °C) has a higher normalized N-to C-terminal contacts than thermophilic GAPDH *Thermus aquaticus*(OGT =70 °C), *Geobacillus Stearothermophilus* (OGT = 65 °C), and mesostable GAPDH from *Escherichia Coli* (OGT = 37 °C) (Table 3). Further, we analyzed N- to C-terminal contacts for other thermophilic/mesophilic protein pairs from different folds (in which N-and C-terminal are in close proximity) and the results are consistent with our findings (data not shown). These observations, taken together affirm the connection between N- to C-terminal non-covalent interactions and protein stability.

It seems interesting that a number of important folds and super folds[26] have their N- and C-termini in contact with each other. Examples drawn from both globular and membrane proteins include, Tata Box Binding Protein-like fold (1PCZ), both the variable and constant domains of immunoglobulin fold (1DBB), β-lactemase fold (1ZG4), Aspartate/ornithine carmoyltransferase like fold (1A1S), rubredoxin-like fold (1BRF), beta-Trefoil (1I1B), Tumor necrosis Factor (TNF)-like fold (1TNF), beta-Grasp (Ubiquitin-like) fold (1UBQ), haloacid dehydrogenase-like fold (1NF2), Ferredoxin-like fold (1VJW), Phosphoglycerate kinase fold (1PHP), armadillo repeat protein (4DB6), Globin-like fold (1BZ0), Lysozyme-like fold (1REX),Thioredoxin fold (1U3I), transmembrane beta-barrels fold (4GCP) and family A G Protein-coupled receptor-like fold (1BRX). 'Making the both ends meet' seems to be a feature common to all these proteins. Proteins might have evolved the N- and C-terminal interactions as one of the strategies to stabilize their structures in a fold specific manner[27]. Thus, it is apparent that in diverse folds/proteins the terminal regions are in close proximity suggesting that they could be considered as candidates for modulating stability by mutation(s) focusing on terminal regions and hence our results should have wider applicability.

Sequence and structure-based bioinformatics analyses have delineated a methodology to identify target positions for mutagenesis that would enhance protein thermostability. In this context, our study reveals that protein termini are one of the regions of interest (ROI) for mutational studies. Mutations focusing on the terminal regions could be considered to modulate the protein stability particularly in folds where terminal regions come together in 3D space, contrary to the general belief that terminal residues are very flexible and hence have less effect on stability[7]. Our work provides the insights into the nature of N- to C-terminal interactions and adds to the repertoire of approaches for increasing thermal stability of proteins. The need to stabilize proteins continues to grow in importance. For example, therapeutic proteins from beta-trefoil fold, Fibroblast growth factor 1 (FGF1) involved in ischemic disease and type-2 diabetes and Fibroblast growth factor 2 (FGF2) involved in wound healing have terminal regions in close proximity. It is reported that the efficacy of these proteins are limited due to low in vivo stability and poor bio-stability[27, 28]. It is rather tempting to suggest that such proteins could be potential candidates for stability enhancement by strengthening N-to C-terminal non-covalent interactions, even though this needs to be established through experimental work. Mutational experiments in future, on both biotechnologically and therapeutically important proteins could benefit from the knowledge that N-C terminal region is also relevant for enhancing protein stability in a fold-specific manner. It is important to investigate more proteins from diverse organisms to decipher other biological significances of N- to C-terminal contacts. Eventually, such studies should help in understanding the evolution and utilization of interactions between termini in the protein universe and for designing effective protein engineering strategies.

**Methods**

**Crystallization and Data Collection.** The purification of recombinant BSX (RBSX) and other N-terminal mutants was carried out as described previously[8]. The RBSX crystals were grown in a reservoir solution containing 0.1M NaCl, 120mM $MgCl_2$, 0.1M Tris-HCl pH 8.5 and 16 % PEG 8000. Similarly, other two mutants V1A and V1L were grown by hanging drop method varying the concentration of $MgCl_2$ and PEG 8000. All the crystals were obtained at 20°C-22°C by setting up protein and reservoir solution in the ratio of 1μl:1μl drop. Home source data sets were collected for RBSX and V1A mutant whereas synchrotron data sets were collected for V1L mutant crystals. All the data sets, RBSX, V1A, and V1L were collected at 100K and data sets were processed using mosflm[29].

**Structure Solution and Refinement.** The structures were solved by the molecular replacement method with program PHASER-MR from the PHENIX program package[30] using the native crystal structure [PDB ID: 2f8q] as the search model. Then Phenix AutoBuild wizard was used for model rebuilding and completion[31]. The AutoBuild Wizard uses RESOLVE, xtriage and phenix.refine to build an atomic model, refine it, and improve the same with iterative density modification, refinement, and model building[32]. Further refinement and model building were carried out using REFMAC5[33, 34] and COOT from the CCP4 program[35]. Five percent of randomly selected observed reflections were kept aside for cross-validation. The stereochemistry of the final models was analyzed with PROCHECK[36] and RMS deviations resulted in the proper values (Table 1). MolProbity[37] was used to validate the final models. None of the models contain residues in the generously and disallowed region of the Ramachandran (φ, ψ) map[38] (Table 1).

**C$_\beta$ Contact order (C$_\beta$CO).** The concept of contact order (CO) was originally used to define the topological complexity of the native protein to explain the differences in folding rates of different protein families[39]:

$$CO = \frac{1}{NL} \sum^{N} |i-j| \, \sigma(r_{i,j}) \qquad (1)$$

Here $\sigma(r_{i,j}) = 1$ when residues i and j are in contact and 0 otherwise. The contact based on any pair of heavy atoms from each residue located within a sphere of threshold distance. L is the length of the protein; N is the total number of contacts within the threshold distance. For the present study, we have defined C$_\beta$ contact order (C$_\beta$CO) as,

$$C_\beta CO = \frac{1}{N} \sum_{\substack{j=1 \\ |i-j|>2}}^{N} |i-j| \sigma(r_{i,j}) \qquad (2)$$

where $\sigma(r_{i,j}) = 1$, if $r_{i,j} < r_d$ and $\sigma(r_{i,j}) = 0$, if $r_{i,j} \geq r_d$, $r_{i,j}$ is the distance exclusively between the $C_\beta$ atoms ($C_\alpha$ atoms for Glycine) of residues i and j in the protein structure. Two residues are considered to be in contact if their $C_\beta$ atoms are closer than a threshold radius $r_d = 10$ Å. Here, L is the length of the protein and N is the total number of $C_\beta$ contacts within the distance threshold. In addition, we have excluded the trivial contacts between the nearest and second nearest residues. Earlier studies in which $C_\beta$ atoms are considered prominently include the definition of half sphere exposure to characterize exposed and buried residues[40], structural alphabets for fold recognition[41] and in the prediction of deleterious mutations[42]. In general, contact order (CO) is used to describe the topological complexity of protein structure and has been reported in connection with protein folding rate and thermostability[39, 43]. The structural metric $C_\beta CO$, used in the current analysis is different from CO. The essence of $C_\beta CO$ is that it considers only the $C_\beta$ atoms instead of all atom models for a residue, while it can retain the intrinsic signature of CO. The main advantage of using only the $C_\beta$ atoms in the calculation is that it is more robust to the poorly defined side chain conformations and simultaneously it has the ability to measure a residue's local side chain environments. This will also be true for those low resolution crystal structures in which side chain density of residues is not well defined and for ab initio model structures or any modelled structures where side chain conformations are not accurately modelled. Moreover, contrary to $C_\alpha$, which is a backbone atom, $C_\beta$ belongs to the side chain atoms of a residue and can provide a side chain centric view of the structure. In addition, $C_\beta CO$ value has a dependency on the direction of a residue's side chain atoms. At the same time, $C_\beta CO$ is easy to compute, conceptually simple to interpret and importantly our works bring out that it correlates with the stability changes of the mutants. Our analysis is presumably among the first studies in which $C_\beta$ atoms are used to study the effect of the long-range order and its implication for protein thermal stability.

**Residue-wise $C_\beta$ contact order (RWC$_\beta$CO).** Kinjo and Nishikawa[44] first introduced the concept of residue-wise contact order, which is the sum of the sequence separation of contacting residues as given in equation (i). For a protein of length L, the RWC$_\beta$CO values of the i-th residue is defined as,

$$RWC_\beta CO_i = \frac{1}{L} \sum_{j:|i-j|>2}^{N} |i-j| \sigma(r_{i,j}) \qquad (3)$$

where $r_{i,j}$ is the distance between the $C_\beta$ atoms of the i-th and j-th residues ($C_\alpha$ for Glycine) in the protein structure as given in equation (ii). We set $r_d = 10$Å as a threshold distance.

**$C_\beta$ contact density ($C_\beta$CD).** We have defined $C_\beta$ contact density as,

$$C_\beta CD = \frac{1}{L} \sum_{i=1}^{L} \sum_{j=1}^{L} n_{i,j} \qquad (4)$$

Where $n_{ij}$ is the total number of $C_\beta$-$C_\beta$ contacts ($C_\alpha$ atoms for Glycine) within a distance cut off (10 Å) and L is the length of the protein structure. We have not used obvious contacts between the nearest and second-nearest residues.

**Structural analysis.** The solvent accessible surface area was calculated using the NACCESS program[45]. Contacts were found using NCONT of the CCP4 suite (CCP4-6.4.0)[46]. Secondary structures were assigned by DSSP[47]. All the figures of molecules were generated using PyMOL (http://www.pymol.org).

**Residue Interaction Networks (RINs).** Protein structure can be represented as a residue interaction network between amino acid residues. In this analysis, protein structures were modelled as undirected graphs in which amino acids are considered as nodes and they are connected by edges or links that correspond to non-covalent interactions between these nodes. These edges of this RIN can be either weighted or un-weighted based on non-covalent interaction strength ($I_s$) between two connecting nodes and/or distances. Here, we used RINerator[48] modules to generate the residue interaction networks for each protein structure. RINerator first adds hydrogen atoms to the original protein structure by using REDUCE[17], and then Probe[18] is used to identify non-covalent residue interactions. Here, the edges are labelled with different interaction types e.g. inter-atomic contacts (cnt), hydrogen bonds (hbond), overlapping van der Waals radii (ovl) and generic residue interactions (combi). Probe also computes interaction score for each edge in which the weight is proportional to the strength of the interaction. It is suggested that in contrast to the other protein network[14, 16], RINerator is capable of generating a more realistic residue interaction networks by sampling atomic packing of each atom using small probe contact dot surface[49].

We constructed different sub networks based on the strength of interaction score ($I_s$) evaluated by Probe between all pairs of amino acid residues in which any pair of amino acid residues is connected by an edge, if their interaction score ($I_s$) is higher than a threshold value ($I_{smin}$). Then, RINs of all three structures were constructed at different $I_{smin}$ values and their network topology and various network parameters were analyzed. All these networks are visualized using Cytoscape[50]. Network Analyzer[51] plugin and RINalyzer[49] plug-in of Cytoscape were used to calculate simple topological parameters and comparison of two RINs respectively.

**Strongly Connected Component (SCC).** Further, we calculated strongly connected component for each network at different cut-off of interaction score by BiNoM2.5[52] module in Cytoscape. BiNoM uses the algorithm of Trojan to decompose the network into strongly connected components[53]. The giant cluster, defined here, as largest strongly connected component (LSCC) is the size of largest group of connected nodes (in terms of number of residues) in the network. The size of the SCC in the network depends on the interaction score cut-off of between two nodes. Hence, the size of SCC is a function of interaction score cut-off ($I_{smin}$). We then calculated the size of largest SCC by varying the cut-off of interaction score and plotted as a function of $I_{smin}$. Here, the size of LSCC is normalized with respect to the total number of residues in the protein.

**Acknowledgements**

P. M. thanks CSIR-UGC, and Indian Institute of Science, Bangalore for financial support. X-ray data for RBSX and V1A crystals were collected at the X-ray Facility for Structural Biology at the Molecular Biophysics Unit, Indian Institute of Science, Bangalore, India. We thank beamline staff at BM14 of the European Synchrotron Radiation Facility (ESRF), Grenoble for providing access to collect the diffraction data of V1L mutant.

**Author Contributions**

P.M., A. B., K. K., V. S. R., and S. R. designed the experiment. P.M., A. B., and K. K., performed the experiment. P. M. crystallized and solved the structures. P.M. and S.R. carried out the computational analysis. All authors wrote and approved the manuscript.

**Competing financial interests**

The authors declare no competing financial interests.

**Figure legends**

**Fig. 1. Overview of RBSX crystal structure**. The crystal structure of RBSX (PDB ID: 4QCE) is shown in cartoon representation. Because of TIM barrel fold, N-terminus (blue region) comes close to the C-terminus (red region) and their proximity is implicated in stability enhancement. The location of mutation is away from the active site (sphere and stick in firebrick) residues (E149 and E259).

**Fig. 2. Overlay of van der Waals contacts at the site of mutation**. Van der Waals contacts between the side chain atoms of mutated residue with surrounding side chain atoms. The dash line represents the van der Waals contact at a distance cut-off of 5Å. Lower panel shows the 2Fo-Fc electron density map of corresponding residues, contoured at 1.0σ level with side chain as sticks.

**Fig. 3. RWC$_\beta$CO comparison of different mutants**. Five-residues moving average of difference of RWC$_\beta$CO between V1L and V1A mutant structures.

**Fig. 4. Comparison of residue interaction network of largest strongly connected component at I$_{smin}$ =1**. Position of unique residues (red sphere) in the three-dimensional structure of V1L and V1A. All residues belong to the cluster of largest strongly connected component (LSCC) when compared between residue interaction network of V1L and V1A at I$_{smin}$=1. Arrows point to terminal residues.

**Fig. 5. Structural superposition of GAPDH from different organisms**. Monomeric unit of (Chain O) of the GAPDH structures is used for alignment (Table 3). Hyperthermophilic GAPDH from *Thermotogamaritima* (1HDG) is taken as reference structure. N-terminal region (blue) and C-terminal region (red) is considered up to second SSE from N-terminal end and from C-terminal end respectively for each structure.

**Table 1**. X-ray Data Collection and Refinement Statistics.

| Name | RBSX | V1A | V1L |
|---|---|---|---|
| Crystallographic data | | | |
|   Space group | $P2_12_12_1$ | C2 | $P2_12_12_1$ |
|   Unit Cell dimensions | | | |
|     a (Å) | 54.77 | 73.57 | 54.88 |
|     b (Å) | 75.65 | 80.12 | 76.58 |
|     c (Å) | 176.91 | 69.90 | 176.73 |
|     α (°) | 90.00 | 90.00 | 90.00 |
|     β (°) | 90.00 | 110.81 | 90.00 |
|     γ (°) | 90.00 | 90.00 | 90.00 |
| Unit Cell Volume (Å$^3$) | 733000.14 | 385141.84 | 742745.00 |
| Data Collection | | | |
| Temperature (K) | 100 | 100 | 100 |
| Resolution (Å) | 27.7-2.32 (2.45-2.32) | 26.8-2.26 (2.38-2.26) | 40.2-1.96 (2.07-1.96) |
| Total Reflections | 30910 | 17095 | 52480 |
| Unique Reflections | | | |
|   Above 1σ | 30459 | 17093 | 52436 |
|   Above 3σ | 21329 | 15556 | 45630 |
|   $R_{merge}$ (%) | 16.1(33.8) | 3.9 (15.3) | 7.6 (21.3) |
|   Average I/σ(I) | 6.7 (3.8) | 25.5 (10.0) | 18.5 (8.7) |
|   Completeness | 95.1 (96.3) | 96.3(94.5) | 97.3(94.3) |
|   Redundancy | 4.7 (4.5) | 5.2 (5.1) | 6.9 (6.6) |
|   Solvent content (%) | 44.97 | 47.55 | 45.14 |
| Refinement Statistics | | | |
| Resolution (Å) | 27.7-2.32 | 26.8-2.26 | 40.2-1.96 |
|   No. of Reflections | 30910 | 17094 | 52480 |
|   $R_{work}$/ $R_{free}$ | 17.9/22.7 | 17.4/22.5 | 15.2/19.0 |
|   No. of atoms | | | |
|     Protein | 5807 | 2905 | 5854 |
|     Ligand/ion | 4 | 5 | 16 |
|     Water | 419 | 202 | 434 |
|   Average B-factors (Å$^2$) | | | |
|     Protein | 21.6 | 26.3 | 13.9 |
|     Ligand/ion | 15.6 | 26.4 | 18.74 |
|     Water | 20.6 | 27.2 | 23.2 |
| RMSD | | | |
|   Bond distance (Å) | 0.007 | 0.005 | 0.019 |
|   Bond angles (°) | 1.123 | 0.955 | 1.802 |
| Luzzati coordinate error(Å) Working set | 0.279 | 0.269 | 0.175 |
| PDB entry | 4QCE | 4QCF | 4QDM |

**Table 2.** $C_\beta CO$ values for RBSX, V1A, and V1L structures. $C_\beta COx$ is calculated for a subset of $C_\beta$ contacts which are unique to V1L excluding V1A contacts and vice-versa. Similarly $C_\beta CDy$ is the unique $C_\beta$ contact density, calculated for a subset of $C_\beta$ contact that is present in V1A but not in V1L and vice-versa. Contacts between the nearest and the second nearest residues are not considered.

| Structure | $C_\beta CO$ | $C_\beta COx$ | $C_\beta CD$ | $C_\beta CDy$ |
|---|---|---|---|---|
| V1A | 52.78 | 53.80 | 7.12 | 0.38 |
| V1L | 53.46 | 75.39 | 7.17 | 0.53 |
| RBSX | 52.93 | | 7.16 | |

**Table 3**. Comparison of N- to C-terminal contacts of GAPDH structures from different organisms.

| PDB | 1HDG | 1CER | 1GD1 | 1GAD |
|---|---|---|---|---|
| Organism | *Thermotoga maritima* | *Thermus aquaticus* | *Geobacillus Stearothermophilus* | *Escherichia Coli* |
| Tm | 80 °C | 70 °C | 65 °C | 37 °C |
| NCn | 4.22 | 3.35 | 3.24 | 2.75 |
| Length | 332 | 333 | 334 | 330 |

Tm = Optimum temperature of the source organism. NCn = Normalized N- to C-terminal contacts

**Figure: 1.**

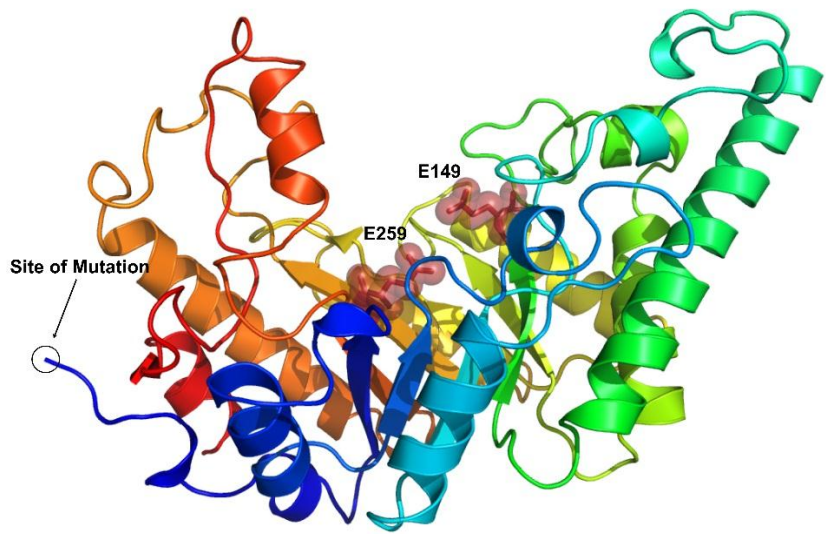

**Figure: 2.**

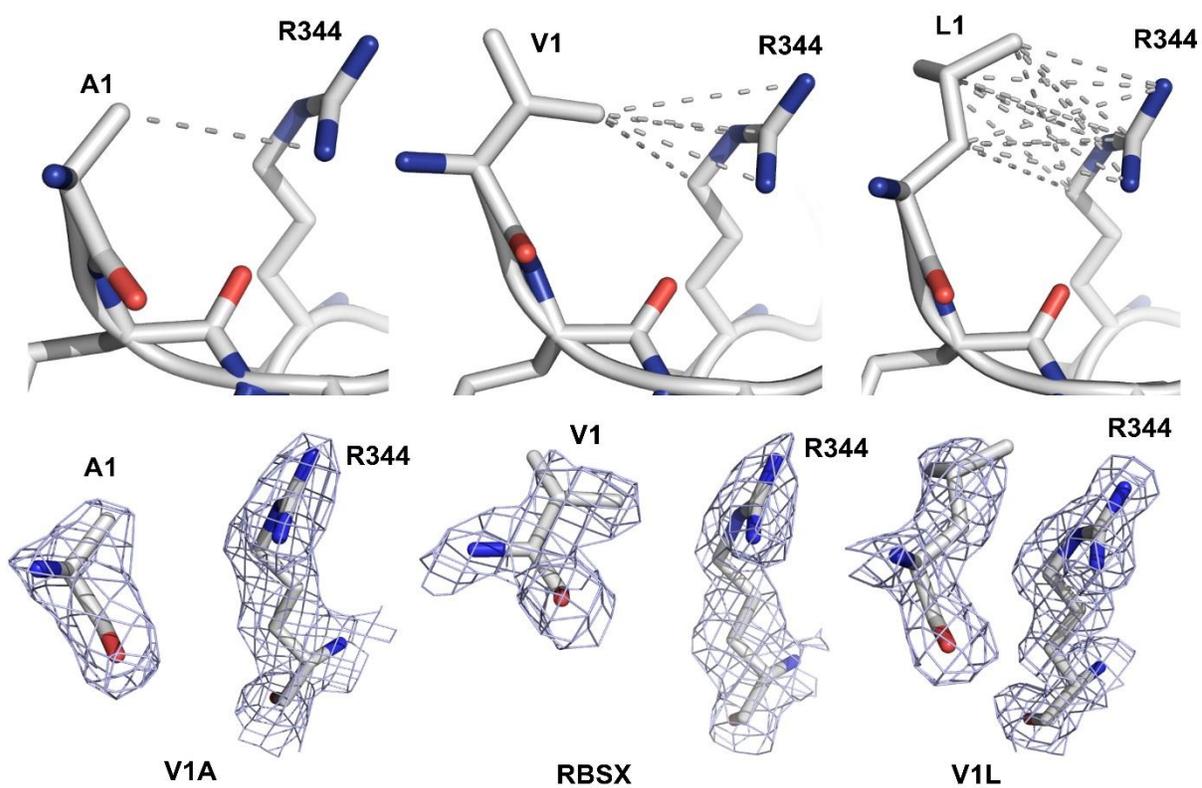

**Figure: 3**.

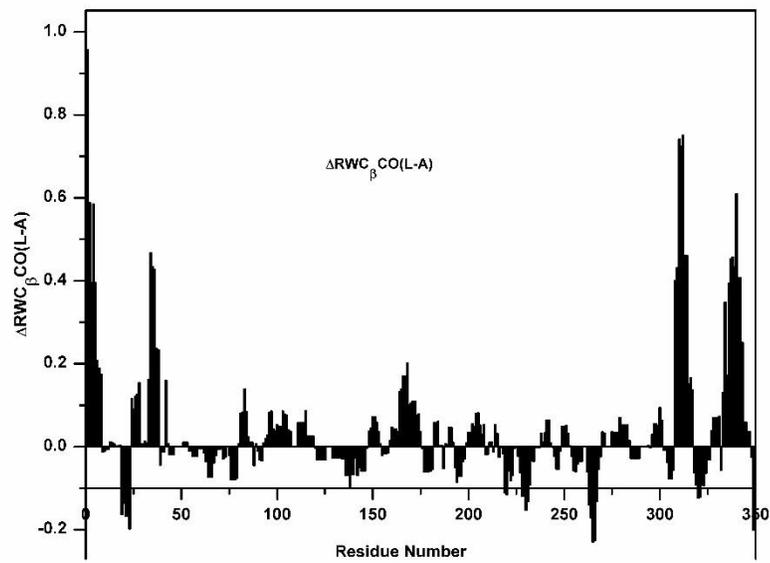

**Figure: 4.**

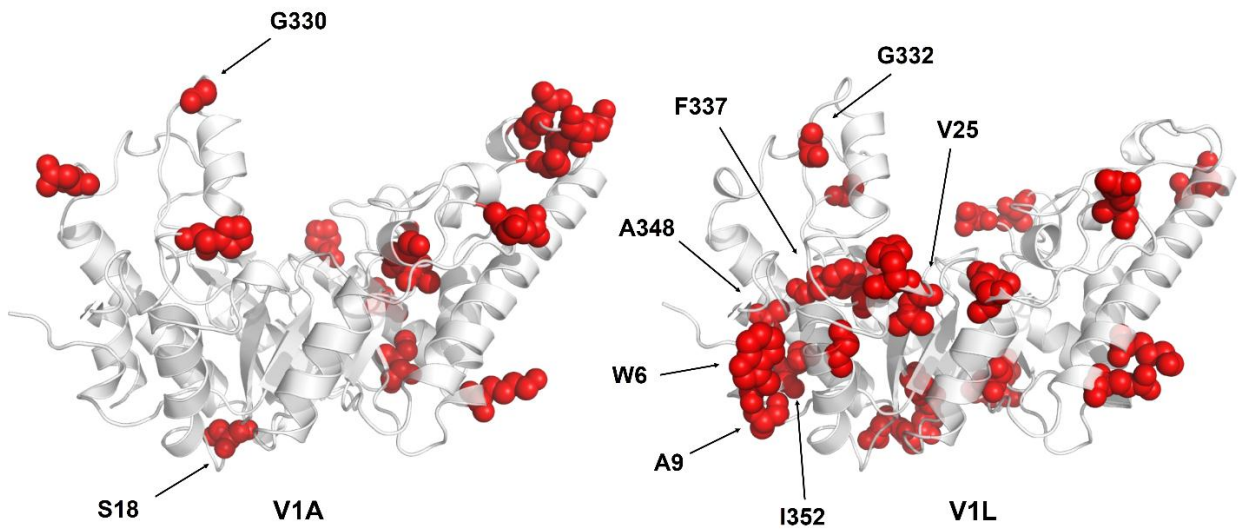

**Figure: 5.**

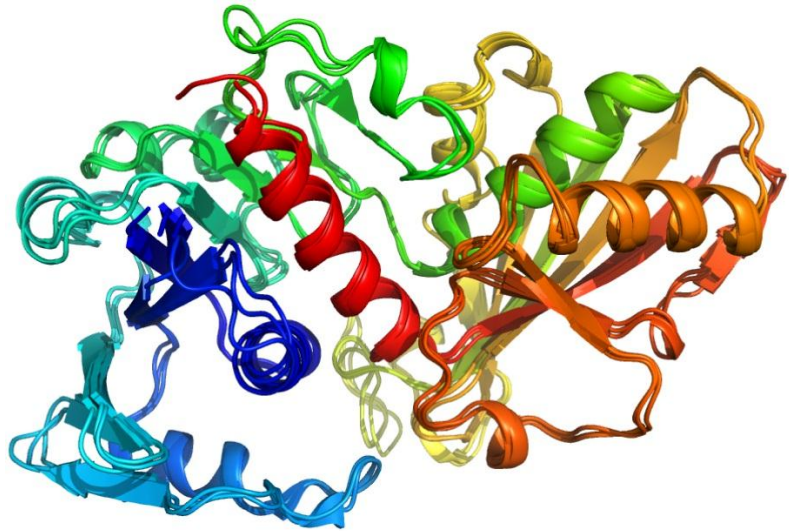

**Modulation of N- to C-terminal interactions enhances protein stability.**

Mahanta P[a], Bhardwaj A[b,c], Kumar K[b], Reddy V S[b], Ramakumar S[a]∗

[a]Department of Physics, Indian Institute of Science, Bangalore-560012; [b]Plant Transformation Group, International Centre for Genetic Engineering and Biotechnology, New Delhi-110067; [c]Present address, Department of Pathology, Kimmel Centre for Biology and Medicine at the Skirball Institute, New York University School of Medicine, New York, 10016, USA

*Corresponding author: ramak@physics.iisc.ernet.in

**Supplementary Information**

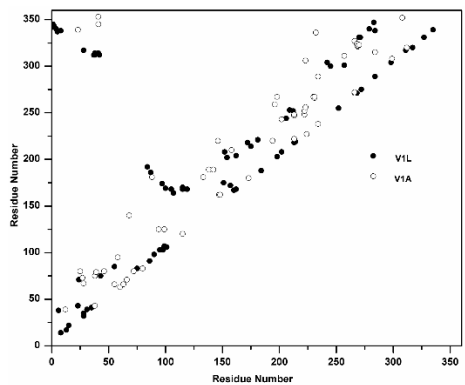

**Fig. S1**. **Difference in C$_\beta$ contacts map.** Difference plot of C$_\beta$ contacts between V1L structure and V1A structure. Here, solid circles and open square are unique C$_\beta$ contacts (cut-off distance, d= 10 Å) corresponds to V1L and V1A respectively.

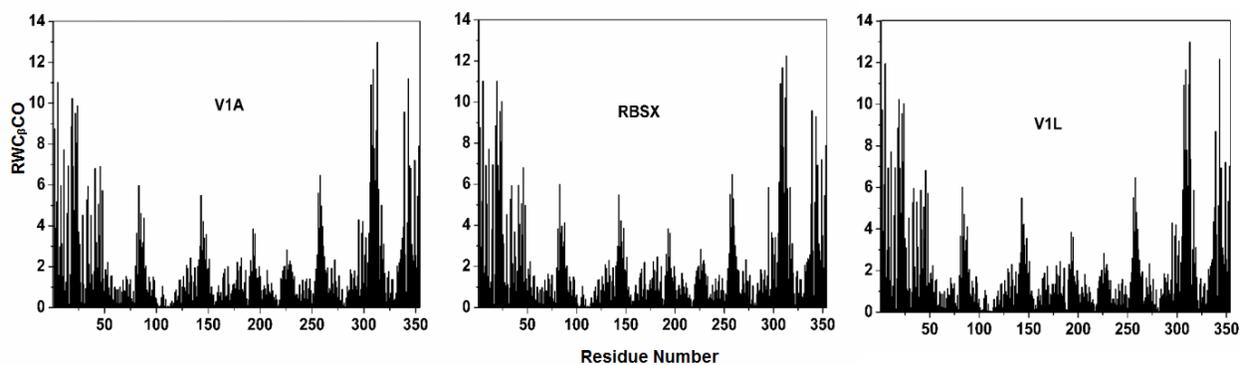

**Fig. S2. RWC$_\beta$CO comparison of different mutants.** Residue wise C$_\beta$ contact order value for each mutant along the primary sequence.

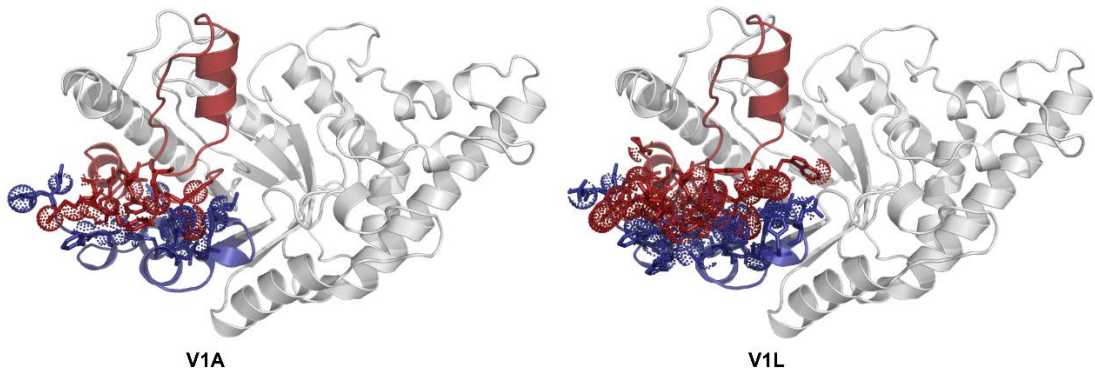

**Fig. S3. Unique N-to C-terminal contacts corresponding to more stable and less stable mutants.** A dot representation of unique contacts between N-terminal segment (deep blue) and C-terminal segments (firebrick)) in V1L mutant and V1A mutant respectively.

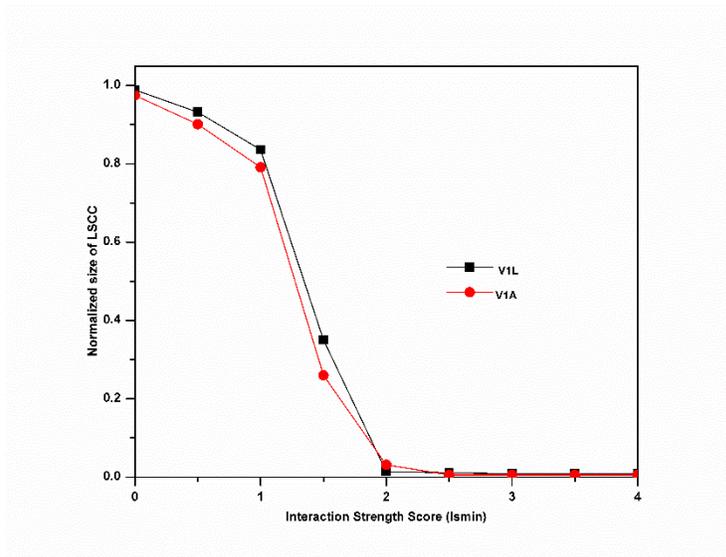

**Fig. S4. Transition profile of largest strongly connected components in the network.** Difference in transition profiles in the size of largest strongest connected component (LSCC) of V1L and V1A mutant structures at different interaction score cut off ($I_{smin}$).

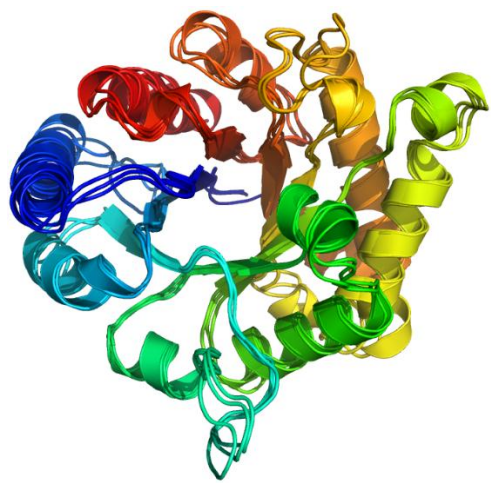

**Fig. S5**. **Structural superposition of Triosephosphate isomerase (TIM) from different organisms.** Monomeric unit (chain A) of TIM (EC 5.3.1.1) across different organisms is used for alignment. Hyperthermophilic TIM from *Thermotoga maritima* (1B9B, OGT= 80 °C) is taken as reference structure. Despite very similar structure (rmsd between hyperthermophilic TIM structure and thermophilic TIM structure (1BTM, OGT= 65 °C) is 0.9568 Å and that between the hyperthermophilic TIM structure and mesophilic TIM structure (1TRE, OGT= 37 °C) is 1.3 Å respectively) they show differences in normalized N- to C-terminal contacts.

**Table S1.** List of van der Waals contacts between the side chain atoms of mutated residue with surrounding side chain atoms at a distance cut-off of 5Å.

| Structure | Atoms of mutated amino acid | Atoms of its surrounding residue | Distances (Å) |
|---|---|---|---|
| **V1A** | 1ALA./CB | 344(ARG)./NH1 | 4.38 |
| **RBSX** | 1VAL./CG1 | 344(ARG)./NE | 3.99 |
| | | 344(ARG)./CZ | 3.84 |
| | | 344(ARG)./NH1 | 3.75 |
| | | 344(ARG)./NH2 | 4.44 |
| | | 344(ARG)./CD | 4.11 |
| **V1L** | 1(LEU)./CB | 344(ARG)./NH2 | 4.81 |
| | | 344(ARG)./CD | 4.72 |
| | | 344(ARG)./NE | 4.61 |
| | | 344(ARG)./CZ | 4.24 |
| | | 344(ARG)./NH1 | 3.84 |
| | 1(LEU)./CG | 344(ARG)./NH2 | 4.76 |
| | | 344(ARG)./NE | 4.74 |
| | | 344(ARG)./CZ | 4.49 |
| | | 344(ARG)./NH1 | 4.50 |
| | 1(LEU)./CD1 | 344(ARG)./NH2 | 3.60 |
| | | 344(ARG)./CD | 4.95 |
| | | 344(ARG)./NE | 4.19 |
| | | 344(ARG)./CZ | 3.71 |
| | | 344(ARG)./NH1 | 3.94 |
| | 1(LEU)./CD2 | 344(ARG)./CD | 4.81 |
| | | 344(ARG)./NE | 4.65 |
| | | 344(ARG)./CZ | 4.83 |

**Table S2.** LRC$_\beta$CO values for RBSX, V1A, and V1L structures at different long-range cut-off.

| Structure | Residue Separation ($\geq$) | LRC$_\beta$CO |
|---|---|---|
| V1A | | 77.61 |
| RBSX | 10 | 77.78 |
| V1L | | 79.12 |
| V1A | | 93.83 |
| RBSX | 30 | 93.89 |
| V1L | | 95.59 |
| V1A | | 126.91 |
| RBSX | 50 | 127.08 |
| V1L | | 129.00 |

**Table S3**. We have used a cut off distance of 5Å to identify a contact between a pair of atom (A): Extra atom-atom contacts between N-terminal segments (Residues 1-25) and C-terminal segment (residues 319-354) in V1L mutant structure with respect to V1A mutant structure.

| Res Num | Res Name | Atom | Res Num | Res Name | Atom | Dist | VdW CC | VdW Others | Polar interactions |
|---|---|---|---|---|---|---|---|---|---|
| 1 | LEU | C | 344 | ARG | NH1 | 4.69 | | 1 | |
| 1 | LEU | CA | 344 | ARG | NH1 | 4.95 | | 1 | |
| 1 | LEU | CB | 344 | ARG | CD | 4.72 | 1 | | |
| 1 | LEU | CB | 344 | ARG | CZ | 4.24 | 1 | | |
| 1 | LEU | CB | 344 | ARG | NE | 4.61 | | 1 | |
| 1 | LEU | CB | 344 | ARG | NH1 | 3.84 | | 1 | |
| 1 | LEU | CB | 344 | ARG | NH2 | 4.81 | | 1 | |
| 1 | LEU | CD1 | 344 | ARG | CD | 4.95 | 1 | | |
| 1 | LEU | CD1 | 344 | ARG | CZ | 3.71 | 1 | | |
| 1 | LEU | CD1 | 344 | ARG | NE | 4.19 | | 1 | |
| 1 | LEU | CD1 | 344 | ARG | NH1 | 3.94 | | 1 | |
| 1 | LEU | CD1 | 344 | ARG | NH2 | 3.6 | | 1 | |
| 1 | LEU | CD2 | 344 | ARG | CD | 4.81 | 1 | | |
| 1 | LEU | CD2 | 344 | ARG | CZ | 4.83 | 1 | | |
| 1 | LEU | CD2 | 344 | ARG | NE | 4.65 | | 1 | |
| 1 | LEU | CG | 344 | ARG | CZ | 4.49 | 1 | | |
| 1 | LEU | CG | 344 | ARG | NE | 4.74 | | 1 | |
| 1 | LEU | CG | 344 | ARG | NH1 | 4.5 | | 1 | |
| 1 | LEU | CG | 344 | ARG | NH2 | 4.76 | | 1 | |
| 2 | GLN | C | 344 | ARG | CZ | 4.97 | 1 | | |
| 2 | GLN | C | 345 | VAL | O | 5 | | 1 | |
| 2 | GLN | CA | 344 | ARG | CB | 4.79 | 1 | | |
| 2 | GLN | CA | 344 | ARG | CD | 4.93 | 1 | | |
| 2 | GLN | CA | 344 | ARG | NH1 | 4.97 | | 1 | |
| 2 | GLN | CB | 344 | ARG | CA | 4.98 | 1 | | |
| 2 | GLN | CB | 345 | VAL | CB | 4.93 | 1 | | |
| 2 | GLN | N | 344 | ARG | CD | 4.65 | | 1 | |
| 2 | GLN | O | 345 | VAL | N | 4.9 | | | 1 |
| 3 | PRO | C | 343 | TYR | O | 4.95 | | 1 | |
| 3 | PRO | CG | 345 | VAL | CG1 | 4.81 | 1 | | |
| 3 | PRO | CG | 350 | TRP | CE2 | 4.97 | 1 | | |
| 3 | PRO | N | 344 | ARG | CB | 5 | | 1 | |
| 3 | PRO | O | 343 | TYR | C | 4.86 | | 1 | |
| 4 | PHE | CD1 | 343 | TYR | CA | 4.71 | 1 | | |
| 4 | PHE | CE1 | 342 | ASN | CB | 4.79 | 1 | | |
| 4 | PHE | CE1 | 343 | TYR | C | 4.81 | 1 | | |
| 4 | PHE | CE1 | 344 | ARG | N | 4.8 | | 1 | |
| 4 | PHE | CG | 343 | TYR | C | 4.94 | 1 | | |

| Res Num | Res Name | Atom | Res Num | Res Name | Atom | Dist | VdW CC | VdW Others | Polar interactions |
|---|---|---|---|---|---|---|---|---|---|
| 4 | PHE | CZ | 344 | ARG | CD | 4.89 | 1 | | |
| 4 | PHE | N | 345 | VAL | N | 4.94 | | | 1 |
| 8 | VAL | CB | 350 | TRP | CZ3 | 4.86 | 1 | | |
| 8 | VAL | CG1 | 345 | VAL | CB | 4.93 | 1 | | |
| 8 | VAL | CG2 | 349 | PHE | CG | 4.71 | 1 | | |
| 9 | ALA | C | 349 | PHE | CE1 | 4.78 | 1 | | |
| 9 | ALA | CA | 349 | PHE | CZ | 4.82 | 1 | | |
| 9 | ALA | N | 349 | PHE | CE2 | 4.94 | | 1 | |
| 9 | ALA | O | 349 | PHE | CD1 | 4.97 | | 1 | |
| 11 | LEU | CD2 | 352 | ILE | CB | 4.96 | 1 | | |
| 14 | ARG | CB | 352 | ILE | O | 4.88 | | 1 | |
| 14 | ARG | CB | 353 | ILE | CA | 4.93 | 1 | | |
| 14 | ARG | CD | 353 | ILE | CG1 | 4.97 | 1 | | |
| 14 | ARG | CD | 354 | ASP | N | 5 | | 1 | |
| 14 | ARG | CG | 352 | ILE | C | 4.81 | 1 | | |
| 14 | ARG | CG | 354 | ASP | O | 4.95 | | 1 | |
| 14 | ARG | NE | 352 | ILE | C | 4.99 | | 1 | |
| 15 | TYR | CE2 | 352 | ILE | CG1 | 4.96 | 1 | | |

**Table S3**. (B): Extra atom-atom contacts between N-terminal segments (Residues 1-25) and C-terminal segment (residues 319-354) in V1A mutant structure with respect to V1L mutant structure.

| Res Num | Res Name | Atom | Res Num | Res Name | Atom | Dist | VdW CC | VdW Others | Polar interactions |
|---|---|---|---|---|---|---|---|---|---|
| 1 | ALA | C | 344 | ARG | NH1 | 4.97 | | 1 | |
| 1 | ALA | CB | 344 | ARG | NH1 | 4.38 | | 1 | |
| 2 | GLN | CD | 346 | LYS | O | 4.98 | | 1 | |
| 2 | GLN | OE1 | 346 | LYS | C | 4.99 | | 1 | |
| 2 | GLN | OE1 | 346 | LYS | O | 4.92 | | | 1 |
| 2 | GLN | OE1 | 350 | TRP | CD2 | 4.99 | | 1 | |
| 2 | GLN | OE1 | 350 | TRP | CG | 4.95 | | 1 | |
| 3 | PRO | CA | 345 | VAL | CG2 | 4.89 | 1 | | |
| 4 | PHE | CD1 | 344 | ARG | NH2 | 4.86 | | 1 | |
| 4 | PHE | CG | 344 | ARG | CZ | 4.88 | 1 | | |
| 5 | ALA | CB | 343 | TYR | CA | 4.89 | 1 | | |
| 5 | ALA | O | 339 | PHE | CE1 | 4.98 | | | 1 |
| 6 | TRP | CH2 | 342 | ASN | O | 4.84 | | 1 | |
| 6 | TRP | CZ3 | 343 | TYR | O | 4.83 | | 1 | |
| 11 | LEU | CD2 | 353 | ILE | CA | 4.94 | 1 | | |
| 11 | LEU | CD2 | 353 | ILE | CB | 4.99 | 1 | | |
| 14 | ARG | CD | 354 | ASP | C | 4.98 | 1 | | |

**Table S4**. Different network parameters of RBSX and its mutant structures at different interaction score ($I_{smin}$) cut-off.

| $I_{smin}$ | 0 | | | 1 | | | 3 | | |
|---|---|---|---|---|---|---|---|---|---|
| | **E** | **Edge/node (E/N)** | **<k>** | **E** | **Edge/node (E/N)** | **<k>** | **E** | **Edge/node (E/N)** | **<k>** |
| V1A | 4587 | 12.96 | 8.525 | 1831 | 5.17 | 4.95 | 341 | 0.96 | 1.85 |
| RBSX | 4594 | 12.98 | 8.554 | 1867 | 5.27 | 4.99 | 363 | 1.02 | 1.90 |
| V1L | 4699 | 13.27 | 8.616 | 1895 | 5.35 | 5.05 | 363 | 1.02 | 1.98 |

**Table S5.** List of unique residues that belongs to LSCC of V1L and V1A at $I_{smin}$=1 respectively. Highlighted residues (bold) are correspond to terminal secondary structural elements (SSE).

| | |
|---|---|
| **V1L** | **W6, A9, V25,** H28, G35, A50, E59, G67, E112, E135, K131, A143, D153, T155, E158, P236, G250, L251, G266, D302, A303, D304, S307, G312, **G332, F337, A348, I352** |
| **V1A** | **S18**, P53, E60, V88, D100, E102, N104, K184, P203, L210, I222, D277, H317, **G330** |